\documentclass[aps,amssymb,amsmath,amsfonts]{revtex4}
\usepackage[dvips]{graphicx}
\usepackage{bm,bbm}

\usepackage[usenames]{color}

\newcommand{\ttbl}[1]{\textcolor{black}{#1}}

\newcommand{\enfasi}[1]{\emph{#1}}
\usepackage{dsfont}

\begin{document}
\title{Random Quantum Operations}

\author{Wojciech Bruzda$^{1}$, Valerio Cappellini$^{{1}}$, Hans-J{\"u}rgen
Sommers$^{{2}}$, and Karol {\.Z}yczkowski$^{1,{3}}$\\[2ex]
 {\normalsize\itshape {$^1$Mark Kac Complex Systems Research Centre, Institute of Physics,
   Jagiellonian University, ul. Reymonta 4, 30-059 Krak{\'o}w, Poland}}\\
 {\normalsize\itshape $^{{2}}$Fachbereich Physik, Universit\"{a}t Duisburg-Essen, Campus Duisburg, 47048 Duisburg, Germany}\\
 {\normalsize\itshape $^{{3}}$Centrum Fizyki Teoretycznej, Polska Akademia Nauk,  Al. Lotnik{\'o}w 32/44, 02-668 Warszawa, Poland}\\[2ex]}

 \date{{October 13, 2008}}

\begin{abstract}
We define a natural ensemble of
 trace preserving, completely positive quantum maps
and present algorithms to generate them at random.
Spectral properties of the superoperator $\Phi$
associated with a given quantum map are investigated
and a quantum analogue of the Frobenius--Perron theorem is proved.
We derive a general formula for the density of eigenvalues of $\Phi$
and show  the connection with the Ginibre
ensemble of real non-symmetric random matrices.
Numerical investigations of the spectral gap imply
that a generic state of the system
iterated several times by a fixed generic
map converges exponentially to an invariant state.
\end{abstract}

\maketitle

\medskip

Random dynamical systems are subject of a considerable scientific
interest. Certain statistical properties of classical deterministic
systems which exhibit chaotic behavior can be described by random
processes. A similar link exists also in the quantum theory, since
quantum analogues of classically chaotic dynamical systems display
spectral properties characteristic of ensembles of random
matrices~\cite{Me91}.

In the case of an isolated quantum system the dynamics is represented
by a unitary evolution operator which acts on the space of pure quantum states.
To characterize such evolution operators for periodically driven,
quantum chaotic systems the  Dyson circular ensembles of random
unitary matrices are often used. The symmetry properties of the
system determine which universality class should be used~\cite{Ha}.

In a more general case of a quantum system coupled with an
environment one needs to work with density operators. In such a case
any discrete dynamics is governed by a \enfasi{quantum
operation}~\cite{BZ06} which maps the set of all density operators
of size $N$ into itself. Investigations of the set ${\cal S_N}$ of
all  quantum operations play the key role in the field of  quantum
information processing~\cite{NC00}, since any physical
transformation of a state carrying quantum information has to be
described by an element of this set.

To process quantum information one needs to
transform a given quantum state in a controlled way.
When designing a sequence of quantum operations, which
constitutes a quantum algorithm, it is important to understand the
properties of each operation and to estimate the
influence of any imperfection in realization
of an operation on  the final result.

In order to describe an effect of external noise acting on a quantum system
one often uses certain models of random quantum operations. On the other hand,
random operations are  purposely applied to obtain pseudo--random quantum
circuits~\cite{ELL05,BWV08}. These possible applications provide motivation
for research on random quantum operations.

The main aim of this work is to construct
a general class of random quantum operations
and to analyze their properties. More formally, we
introduce  a natural probability measure which covers the entire set
${\cal S}_N$ of quantum operations
and present an efficient algorithm to generate them randomly.
We investigate spectral properties of superoperators
associated with quantum operations and infer
conclusions about the convergence of any initial state
subjected to the repeated action of a given random
operation to its invariant state.
It is worth to emphasize
that the spectral properties of superoperators
are already accessible experimentally,
as demonstrated by Weinstein et al.~\cite{WHO04}
in a study of a three-qubit nuclear
magnetic resonance quantum information processor.

We start reviewing the classical problem, in which classical
information is encoded into a probability vector $\vec p$ of length
$N$. The set of all classical states of size $N$ forms the
probability simplex $\Delta_{N-1}$. A discrete dynamics in this set
is given by a transformation $p^\prime_i=S_{ij}p_j$, where $S$ is a
real square matrix of size $N$, which is \enfasi{stochastic},
i.e.:

\noindent a) $S_{ij}\ge 0$ for $i,j=1,\dots, N$;

\noindent
b) $\sum_{i=1}^N S_{ij}\ = \  1\ $ for all $j=1,\dots, N$.
\smallskip
 
\noindent {\sl If a real matrix $S$ satisfies assumptions (a) and
(b) then

\noindent
i) the spectrum $\{ z_i\}_{i=1}^N$ of $S$ belongs to the unit disk,
$|z_i|\le 1$,  and
the leading eigenvalue equals unity, $z_1=1$;

\noindent ii) \ttbl{the eigenspace associated with $z_1$ contains
 (at least) a
real eigenstate ${\vec p}_{\rm inv}$, which describes the invariant
measure of $S$}.
\smallskip

\noindent
If additionally

\noindent
c) $\sum_{j=1}^N S_{ij}\ = \  1\ $ for all $i=1,\dots, N$

then the matrix $S$ is called \enfasi{bistochastic} (doubly stochastic)
 and

\noindent
iii) the maximally mixed state is invariant,
${\vec p}_{\rm inv} = (1/N,\dots, 1/N)$.
}

This is a form of the well-known {\bf Frobenius--Perron (FP)
theorem} and its proof may be found e.g. in~\cite{MO79,Ber05:1}.

To generate a random stochastic matrix it is convenient to start with
a  square matrix $X$ from the complex Ginibre ensemble, all elements of
which
are independent  complex random Gaussian  variables.
Then the random matrix
\begin{equation}
S_{ij} \ := \ |X_{ij}|^2/ \sum_{m=1}^N |X_{mj}|^2 \ ,
\label{randstoch}
\end{equation}
is stochastic~\cite{Zyk01:1}, and each of its columns forms an
independent random vector distributed uniformly in the probability
simplex $\Delta_{N-1}$.

Let us now discuss the quantum case, in which \ttbl{any state can be
described by means of} a positive, normalized operator,
 $\rho\ge 0$, Tr$\rho=1$. Let ${\cal M}_N$ denote
the set of all normalized states which act on a $N$-dimensional
Hilbert space ${\cal H}_N$.
A quantum analogue of a stochastic
matrix is given by a linear quantum map $\Phi: {\cal M}_N \to {\cal
M}_N$ which preserves the trace, and is completely positive (i.e.
the extended map, $\Phi\otimes \ttbl{\mathds{1}_K}$,
 is positive for any size $\ttbl{K}$ of the extension $\ttbl{\mathds{1}_K}$).
Such a transformation \ttbl{is} called \enfasi{quantum operation} or
\enfasi{stochastic map}, \ttbl{and} can be described by a matrix
$\Phi$ of size $N^2$,
\begin{equation}
\rho^\prime\: = \:  \Phi \rho \quad \quad {\rm or} \quad \quad
\rho_{m\mu}\!\!\!\!\!\!^\prime\ \ \: = \:
 \Phi_{\stackrel{\scriptstyle m \mu}{n \nu}}
  \, \rho_{n \nu} \ .
 \label{dynmatr1}
\end{equation}
It is convenient to reorder elements of this
matrix~\cite{SMR61,BZ06} defining the \ttbl{so-called}
\enfasi{dynamical matrix} $D$,
\begin{equation}
D ({\Phi}) \: \equiv \:  \Phi^R
 \quad {\rm so \quad that }\quad
D_{\stackrel{\scriptstyle m n}{\mu \nu}} =
\Phi_{\stackrel{\scriptstyle m \mu }{n \nu}} \ ,
 \label{dynmatr3}
\end{equation}
since the matrix $D$ is Hermitian if the map $\Phi$ preserves
Hermiticity. Note that the above equation can be interpreted as a
definition of the operation of \enfasi{reshuffling} (realignment) of
a matrix $X$, denoted by $X^R$. This definition is representation
dependent, and corresponds to exchanging the position of some
elements of a matrix. Obviously reshuffling is an involution:
$D^R=\Phi$.

Due to the theorem of Choi~\cite{Cho75a}, a map $\Phi$ is
completely positive (CP) if and only if  the corresponding dynamical matrix $D$ is
positive, $D\ge 0$. Hence $D$ can be interpreted as a positive operator
acting on a composed Hilbert space ${\cal H}:={\cal H}_A \otimes {\cal H}_B$.
Any completely positive map can \ttbl{also} be
written in the \enfasi{Kraus form}~\cite{Kr71},
\begin{equation} \rho \: \rightarrow \:
\rho^\prime \: = \:  \sum_{i} A_i\rho A^{\dagger}_i \ .
\label{Kraus1}
\end{equation}
The set of Kraus operators $A_i$ allows one to write down the
linear superoperator as
\begin{equation}
\Phi\: =\: \sum_{i}  A_i \otimes {\bar A}_i \ ,
\label{dynmatr5c}
\end{equation}
where  ${\bar A}_i$ is the  complex conjugate of $A_i$,
while $A_i^{\dagger}$ denotes the adjoint operator:
$A_i^{\dagger}={\bar A_i}^T$.

The map $\Phi$ is \enfasi{trace preserving}, ${\rm Tr}
\rho^\prime={\rm Tr}\rho=1$,
 if  $\sum_iA^{\dagger}_iA_i  =  \mathds{1}_N$.
This condition is equivalent
to a partial trace condition imposed on the dynamical
matrix,
\begin{equation}
{\rm Tr}_A D({\Phi})  =  \mathds{1}_N \ ,
\label{partrace}
\end{equation}
which implies Tr\ $D=N$.

Since the dynamical map of an operation $\Phi$ is positive and
normalized{,} the rescaled matrix $D/N$ may be considered as a state
in an extended Hilbert space ${\cal H}_A \otimes {\cal H}_B$ of size
$N^2$. {Stochastic maps and states on the extended space ${\cal H}_A
\otimes {\cal H}_B$ are related by the so--called
\enfasi{Jamio{\l}kowski} isomorphism~\cite{Ja72,BZ06}\,. Making use
of the maximally entangled bipartite state
$|\psi_+\rangle=\frac{1}{\sqrt{N}} \sum_{i=1}^N | i,i\rangle$\,,
this can be expressed as $D({\Phi})/N=(\Phi \otimes \mathds{1})|\psi_+\rangle \langle \psi_+|$\,.} 
For completeness we sketch here a compact proof of the  latter fact. \\
\noindent {\bf Proof}. The projector 
$|\psi_+\rangle \langle \psi_+|$ can be recasted in the form 
$\frac{1}{N} \sum_{i,j=1}^N |i\rangle\langle j |_A\otimes
|i\rangle\langle j |_B$\,, in which the matrix elements on ${\cal
  H}_A$ and ${\cal H}_B$ are explicit. Then, 
$(\Phi_A \otimes \mathds{1}_B)|\psi_+\rangle \langle
\psi_+|=\frac{1}{N} \sum_{i,j=1}^N \Phi\left(|i\rangle\langle j |\right)_A\otimes
\left(|i\rangle\langle j |\right)_B$\,, which in turn can be expressed
in coordinates as
$\frac{1}{N} \sum_{i,j,\mu,\nu=1}^N
\Phi_{\stackrel{\scriptstyle \mu \nu}{i j}}
\:|\mu i\rangle\langle \nu j |$\,. Thus the result follows from
definition~\eqref{dynmatr3}\,.\hfill \rule[0pt]{7pt}{7pt}\\

A quantum map is called \enfasi{unital} if it leaves the maximally
mixed state invariant. It is so if the Kraus operators satisfy
$\sum_i A_i A^{\dagger}_i  =  \mathds{1}_N$. The unitality
condition may also be written in a form ${\rm Tr}_B D  =  \mathds{1}_N$, dual to~\eqref{partrace}\,. A CP quantum map which is trace
preserving and unital is called \enfasi{bistochastic}.

Spectral properties of positive operators
were studied in the mathematical literature~\cite{EHK78,Gr82}
for a general framework of $C^*$-algebras.
Here we analyze spectral properties of the operator $\Phi$
corresponding to a stochastic map
and formulate a  \enfasi{{\bf quantum analogue of  the Frobenius--Perron theorem}}.

\noindent
{\sl 
Let $\Phi$ be a complex square matrix of size $N^2$,
so that it represents an operator
acting in a composite Hilbert space
${\cal H}_{N^2}
={\cal H}_{A} \otimes {\cal H}_{B}$.
Let us order its complex eigenvalues according to their moduli,
$|z_1|\ge |z_2|\ge \dots\ge |z_{N^2}|\ge 0$.

\noindent
Assume that $\Phi$ represents a stochastic quantum map, i.e.
it satisfies

\noindent a$^\prime$) $\Phi^R \ge 0$;
 \quad b$^\prime$)
$ \sum_k \Phi_{\stackrel{\scriptstyle kk}{ij}}=\delta_{ij}$ \
so that
\  ${\rm Tr}_A \Phi^R=\mathds{1}$,

\noindent where the reshuffling operation, denoted by  $^R$, is
defined in eq.~\eqref{dynmatr3}\,.
Then

\noindent i$^\prime$) the spectrum $\{ z_i \}_{i=1}^{N^2}$ of $\Phi$
belongs to the unit disk,  $|z_i|\le 1$, and
 the leading eigenvalue equals unity, $z_1=1$,

\noindent ii$^\prime$) \ttbl{one of the corresponding eigenstates
     forms a matrix $\omega$ of size $N$ which is positive,
     normalized ($\mathrm{Tr} \ \omega=1$) and
     is invariant under the action of the map,
     $\Phi(\omega)=\omega$.}
\smallskip

\noindent If additionally

\noindent c$^\prime$) ${\rm Tr}_B \Phi^R=\mathds{1}$,
\noindent
 then the map is called bistochastic and

\noindent iii$^\prime$) the maximally mixed state is invariant,
    $\omega=\mathds{1}/N$.
}
\smallskip

\noindent {\bf Proof}. Assumption (a$^\prime$) implies that the
quantum map is completely positive while (b$^\prime$) implies the
trace preserving property. Hence $\Phi$ is a linear map which sends
the convex compact set ${\cal M}_N$ of mixed states into itself. Due
to the Schauder fixed--point theorem~\cite{Sch30:1} such a
transformation has a fixed point---an invariant state  $\omega\ge 0$
such that $\Phi \omega=\omega$. Thus $z_1=1$ and all eigenvalues
fulfil $|z_i|\le 1$,
 since otherwise the
assumption that $\Phi$ maps the compact set ${\cal M}_N$ into itself
would be violated. \hfill$\square$

The spectral properties of stochastic maps are similar to those of
classical stochastic matrices discussed in~\cite{ZKSS03}. As noted
in~\cite{PH93,TDV00} the spectrum of the superoperator
is symmetric with respect to the real axis -- see Fig. 1.
Such spectra for a quantum map corresponding to
a classically chaotic irreversible system where studied in~\cite{LPZ02}.

\begin{figure}[htbp]
\centering
\includegraphics[width=0.52\textwidth]{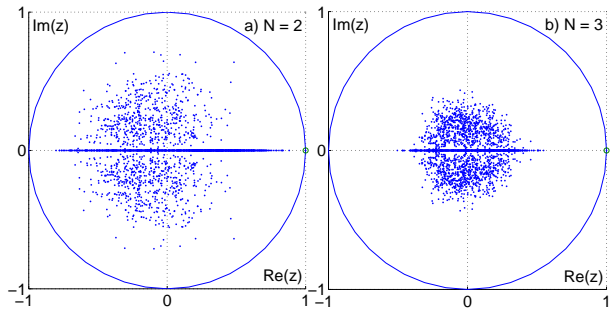}
\caption{Spectra of an ensemble of random superoperators $\Phi$:
a)  800 operations for $N=2$ and b) 300 operations for $N=3$.}
\label{fig:fig1}
\end{figure}

Let us analyze briefly some particular cases of quantum operations.
For any unitary rotation, $\Phi(\rho)=U\rho U^{\dagger}$ the  Kraus
form~\eqref{Kraus1} consists of a single operator only, $A_1=U$.
Thus according to~\eqref{dynmatr5c} the superoperator is given by a
unitary matrix
 $\Phi  =  U \otimes {\bar U}$ of size $N^2$.
Denote the eigenphases of $U$ by $\alpha_i$ where $i=1,\dots N$.
Then the spectrum of $\Phi$ consists of $N^2$ phases given by
$\alpha_i- \alpha_j$ for $i,j=1,\dots,N$. All diagonal phases for
$i=j$ are equal to zero, hence the leading eigenvalue of the
superoperator $\Phi$\,, $z_1=1$\,, exhibits multiplicity
 not smaller than $N$.

Consider now a quantum map for
 which  $\Phi^R$ is diagonal.
This special case can be treated as \enfasi{classical}, since then
$\Phi$ describes a classical dynamics in $\Delta_{N-1}$, while  the
generalized quantum version of the FP theorem reduces to its
standard version. Reshaping \ttbl{a} diagonal dynamical matrix of
size $N^2$ one obtains then \ttbl{a} matrix $S$ of size $N$, where
$S_{ij}=\Phi_{\stackrel{\scriptstyle i j}{i j }}$,
 (no summation performed!).
Then assumption (a$^\prime$) (all diagonal elements of $\Phi^R$ are
positive) gives (a), while the trace preserving condition
(b$^\prime$) implies the probability preserving condition (b).
Similarly, the additional condition (c$^\prime$) for quantum
unitality gives condition (c) which imposes that the uniform vector
is invariant under multiplication by a bistochastic matrix $S$.
Similarly, conclusions (i$^\prime$), (ii$^\prime$) and
(iii$^\prime$) of the quantum version of the theorem imply
conclusions (i--iii) of the standard (classical) Frobenius--Perron
theorem.

We are interested in defining an ensemble of random
operations~\cite{BZ06}. A simple choice of \enfasi{random external
fields}~\cite{AL87},
defined as a convex combination of an arbitrary number
$k$ of unitary transformations,
$\rho^\prime= \sum_{i=1}^k p_i V_i \rho V_i^{\dagger}$, produces
bistochastic operations only.

Let us then consider first a method of
constructing random states by $N \times M$ rectangular random complex
matrices of the Ginibre ensemble~\cite{Gi65}. Taking
\begin{equation}
\rho \ := \ XX^{\dagger}/ {\rm Tr} XX^{\dagger}  \ ,
\label{rhoHS}
\end{equation}
we get a positive normalized state $\rho$. For $M=1$ we obtain a
recipe to generate random pure states, while for $M=N$ the measure
induced by the Ginibre ensemble coincides with the Hilbert--Schmidt
measure in ${\cal M}_N$~\cite{SZ04} and the average purity, $\langle
{\rm Tr} \rho^2\rangle_{\rm HS}$, scales as
$1/N$~\cite{Zyk01:1,BZ06}.

Here we propose an analogous algorithm of
constructing a random operation:

\noindent
1) fix $M \ge 1$ and take a $N^2 \times M$ random complex
Ginibre matrix $X$;

\noindent
2) find the positive matrix $Y:={\rm Tr_A} XX^{\dagger}$
    and its square root $\sqrt{Y}$;

\noindent 3) write the dynamical matrix  (\enfasi{Choi matrix})
\begin{equation}
D=
\Bigl(\mathds{1}_N \otimes \frac{1}{\sqrt Y} \Bigr)
 XX^{\dagger}
\Bigl( \mathds{1}_N \otimes \frac{1}{\sqrt Y}\Bigr)
 \ ;
\label{Dren}
\end{equation}

\noindent 4) reshuffle the Choi matrix according to~\eqref{dynmatr3}
to obtain the superoperator $\Phi=D^R$, and use it as
in~\eqref{dynmatr1} to produce a random map.

It is not difficult to check that  the relation~\eqref{partrace}
holds due to~\eqref{Dren}\,,
 so the random map
preserves the trace. Such a renormalization
to obtain the Choi matrix was independently used in~\cite{AS07}.
This method is simple to apply
for numerical simulations, and exemplary spectra
obtained in the case $M=N^2$ are shown in Fig.1.

For larger $N$, the subleading eigenvalue modulus
$r=|z_2|$ is smaller, so the convergence rate
of any initial $\rho_0$ to the invariant state $\omega$
occurs faster. To demonstrate this effect
we studied the decrease of an average
trace distance in time,
$L(t)=\langle {\rm Tr} |\Phi^t(\rho_0)-\omega|\rangle_{\psi}$,
where the average is performed over an ensemble
of initially pure random states, $\rho_0=|\psi\rangle\langle\psi|$.
Numerical results confirm an exponential convergence,
$L(t)\sim \exp(- \alpha t)$.
The mean convergence rate $\langle \alpha \rangle_ {\Phi}$,
 averaged over an ensemble of random operations,
increases with the dimension $N$ like $\log N$, with slope
very close to unity -- see Fig. 2.
\begin{figure}[htbp]
\centering
\includegraphics[width=0.46\textwidth]{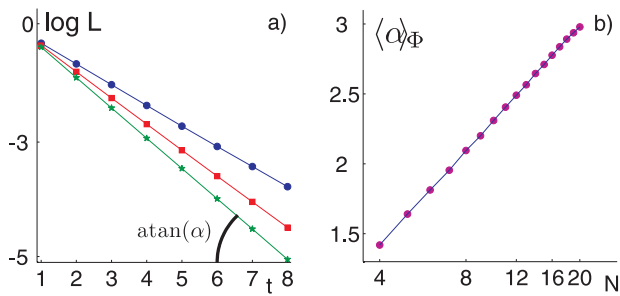}
\caption{
 a) Average trace distance of random pure states to the invariant state
of $\Phi$ as a function of time for \mbox{$N=4  (\bullet), \; 6
(\blacksquare),  \; 8 (\star)$;}
b) mean convergence rate  $\langle \alpha \rangle_ {\Phi}$ as a function
of the system size $N$,  plotted in log scale.
}
\label{fig:fig2}
\end{figure}

To explain these findings we need to analyze
 spectral properties of an ensemble of random operations.
Consider first a random Choi matrix $D$ obtained by the above algorithm
from a Ginibre matrix $X$,
 generated  according to the distribution
  $\propto\exp(-{\rm  Tr} XX^{\dagger})$.
Then $D$  is of the Wishart type and has the distribution
\begin{equation}
 P(D)\propto \int dX \exp(-{\rm Tr} XX^{\dagger})
\delta(D- W XX^{\dagger}W^{\dagger}) \ ,
\end{equation}
where $W:= \mathds{1} \otimes ({\rm Tr}_A XX^{\dagger})^{-1/2}$.
This integral can be rewritten  with help
of  another Delta function,
\begin{equation}
 P(D) \ \propto \
\int \!\! dY \!\! \int \!\! dX \exp(-{\rm Tr} Y)
\delta(D- WXX^{\dagger}W^{\dagger}
) \delta(Y- {\rm Tr}_A XX^{\dagger}) \ .
\end{equation}
Using the Delta function property
and taking the Jacobians into account  we arrive at
\begin{equation}
 P(D)\ \propto \  {\rm det} (D^{M-N^2})\;
\delta({\rm Tr_A}D\ -\mathds{1})  \ ,
\label{Ddist}
\end{equation}
which shows that there are no other constraints on the distribution
of the Choi matrix, besides the partial trace condition and
positivity. This is equivalent to saying that the $M$ Kraus matrices
$A_i$, which form a map $\Phi$,
 constitute a  $(MN) \times N$ truncated part of a unitary matrix $U$
of size $NM$. A natural assumption is that the matrix  $U$ is
distributed according to the Haar measure. Unitarity constraints
become weak  for large $N$,  so the non--Hermitian $N\times N$
truncations $A_i$ are described by the complex Ginibre ensemble~\cite{Gi65}:
Their spectra cover uniformly the disk of radius $1/\sqrt{N}$ in the
complex plane~\cite{ZS00}.

Therefore we are in position to present an alternative
algorithm of generating the same ensemble of random operations,
which has a simple physical interpretation:

\noindent (1$^\prime$) Choose a random unitary matrix $U$ according
to the Haar measure
    on $U(NM)$;

\noindent (2$^\prime$) Construct a random map defined by
\begin{equation}
\rho^\prime \ = \  {\rm Tr}_M [U(\rho \otimes |\nu\rangle\langle
\nu|)U^{\dagger}] \ .
\label{envU}
\end{equation}
\noindent
Hence this random operation corresponds to an interaction with
 an  $M$--dimensional environment,
 initially  in a random  pure state $|\nu \rangle$.
Of a special importance is the case $M=N^2$, for which the term with
the determinant in~\eqref{Ddist} disappears, so the
 matrices $D$ are generated according to the
measure analogous to the
Hilbert-Schmidt measure. This case
provides thus generic dynamical matrices of a full rank
and can  be recommended for numerical implementation.
\smallskip

To analyze spectral properties of a superoperator $\Phi$
let us use the Bloch representation of a state $\rho$,
\begin{equation}
\rho=\sum_{i = 0}^{N^2-1} \tau_i \; \lambda^i \ ,
\label{eq2010}
\end{equation}
where $\lambda^i$ are generators of $\textsf{SU(N)}$ such that
$\textsf{tr}\left(\lambda^i\lambda^j\right)=\delta^{ij}$ and
$\lambda^0 =\mathds{1}/ \sqrt{N}$. Since $\rho=\rho^{\dagger}$,
the generalized Bloch vector
$\overrightarrow{\tau}=[\tau_0,\dots,\tau_{N^2-1}]$\ttbl{, also
called \enfasi{coherence vector},} is real. Thus the action of  the
map $\Phi$ can be represented as
\begin{equation}
\tau^\prime=\Phi(\tau)=C\tau+\kappa,
\end{equation}
where $C$ is a real asymmetric contraction matrix of size $N^2-1$
while $\vec \kappa$ is a translation vector,
which vanishes for bistochastic maps.
Their elements can be expressed in terms of the
Kraus operators, e.g.
$C_{ij} ={\rm Tr} \sum_k\lambda^i    A_k  \lambda^j  A_k^{\dagger}$,
while
$\kappa_i =  {\rm Tr} \lambda^i \lambda^0\sum_k \tau_k  A_k^{\dagger}$.
Thus there exists a real  representation of the superoperator
\begin{equation}
\Phi \  =\  \left[
\begin{array}{ll}
  1 & 0 \\
  \kappa & $C$ \\
\end{array}
\right] \ .
\end{equation}
Eigenvalues of $C$, denoted by
 $\{\Lambda_i\}_{i=1}^{N^2-1}$,
 are also eigenvalues of $\Phi$.
We are going to study the case $M=N^2$, for which the
distribution~\eqref{Ddist} simplifies. Like for the real Ginibre
ensemble~\cite{Lehman,SW08} one may derive in this case the measure in
the space of eigenvalues $\Lambda_i$ of $C$
\begin{equation}
d\mu(\Lambda) \ = \ |\prod_i d \Lambda_i| \;
\prod_{k<l} |\Lambda_k - \Lambda_l| \; G(\Lambda) \; ,
\label{mu1}
\end{equation}
where $G(\Lambda)$ is given by the distribution of (real)
traces,
\begin{equation}
 G(\Lambda) \ := \Bigl\langle
 \prod_{\nu=1}^{N^2-1} \delta\Bigl(
\sum_{i=1}^{N^2-1} (\Lambda_i)^{\nu} - {\rm Tr}C^{\nu}
 \Bigr) \Bigr\rangle
\label{mu2}
\end{equation}
with the average
\begin{equation}
 \langle f(C)  \rangle  \ := \
  \int dC d\kappa \; \Theta(D \geqslant 0) \; f(C) \ .
\label{mu3}
\end{equation}
The domain of integration is given by the conditions for complete
positivity, $D\ge 0$, which is not easy to work with, even for
$N=2$. For large $N$ we can expect that these conditions do not play
an important role, so  the dependence $G(\Lambda)$ is weak, and the
measure for $C$ can be described  by the real Ginibre ensemble of
non-\ttbl{symmetric} Gaussian matrices. The spectrum of such random
matrices consists of a component on the real axis, the probability
density of which is given asymptotically by the step function
 $P(x)=\frac{1}{2} \Theta(1-|x|)$~\cite{So07,FN07},
while remaining eigenvalues cover uniformly the unit circle
according to the \enfasi{Girko distribution}~\cite{Me91}.

To analyze  the spectra of random
operators $\Phi$ one needs to set the scale.
 The mean purity
of a random state $\sigma$ of size $N^2$ behaves as
$N^{-2}$~\cite{Zyk01:1} and $D=N\sigma$, thus the average ${\rm
Tr}D^2 = {\rm Tr}\Phi\Phi^{\dagger}$
 is of the order of unity. Hence,
the rescaled matrix $\Phi^\prime:=N\Phi$  of size $K=N^2$ has the
normalization ${\rm Tr}\Phi^\prime(\Phi^\prime)^{\dagger} \approx K$, which
assures that the radius of the circle is equal to unity.
\begin{figure} 
\centering
\includegraphics[width=0.49\textwidth]{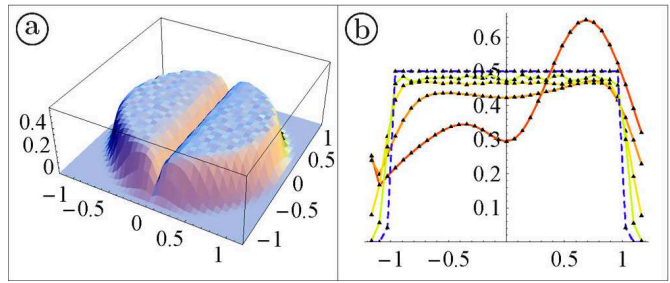}
\caption{ a) Distribution of complex eigenvalues of $10^4$ rescaled
random operators $\Phi^\prime$ already for $N=10$ can be
approximated by the circle law. b) Distribution of real eigenvalues
$P(x)$ of $\Phi^\prime$ plotted for $N=2,3,7$ and $14$ tends to the
step function, characteristic of real Ginibre ensemble. }
\label{fig:fig3}
\end{figure}

Thus we arrive at the following conjecture:
for large $N$ the
statistical properties
of a rescaled random superoperator $N\Phi$ are
described by the real Ginibre ensemble. We confirmed this conjecture
by a detailed numerical investigation. Fig. 3 shows the density of
complex eigenvalues of random superoperators for $N=10$
with $M=N^2$
and the distribution $P(x)$ of the real eigenvalues. As the spectrum of the
rescaled operator $\Phi^\prime=N\Phi$ tends to be localized in the
unit circle, we infer that the size of the subleading eigenvalue
$r=|z_2|$ of $\Phi$ behaves as $1/N$,
hence the convergence rate $\alpha$ scales as $\ln N$.

Numerical studies were also performed for random maps
acting on states of a fixed dimension $N$.
In this case the subleading eigenvalue of a random map~\eqref{envU}
decreases with the varying size of the environment $M$  as $r\sim 1/\sqrt{M}$.
Similar investigations were also performed under a constraint 
that the dynamical matrix  $D$ is diagonal. In this case the
assumption $M=1$
allows to obtain a  random stochastic matrix $S$ of size $N$,
such that each of its columns is generated independently 
with respect to the flat measure in the $(N-1)$
dimensional simplex of probability distributions.
Analyzing the average trace of $SS^T$ 
we infer that in this case the complex spectrum 
can be described by the Girko distribution, which 
covers uniformly the disk of  radius $r\sim 1/\sqrt{N}$.
These spectral properties of random stochastic matrices,
confirmed by our numerical results,
were rigorously analyzed in a recent paper of Horvat
\cite{Ho08}.

In this work we analyzed superoperators associated with
quantum stochastic maps and their spectral properties
and formulated a quantum analogue of the Frobenius--Perron theorem.
We defined an ensemble of random operations,
presented an explicit algorithm to generate them,
and showed an exponential convergence
of a  generic state of the system to the invariant state
under subsequent action of a fixed map.
We demonstrated that for a large dimension of the Hilbert space,
used to describe quantum dynamics,
the spectral properties of a  generic superoperator
can be described by the Ginibre ensemble
of real random matrices.

It is a pleasure to thank M.D. Choi, W.~S{\l}omczy{\'n}ski
and J. Zakrzewski for helpful remarks.
We acknowledge financial support by the SFB Transregio-12 project,
the European grant COCOS,
and a grant number N202 099 31/0746 of Polish Ministry of Science
and Education.

\end{document}